\documentstyle[12pt]{article}
\newcommand{\Frac}[2]{\frac{\displaystyle #1}{\displaystyle #2}}

\begin{document}
\pagestyle{empty}
\begin{titlepage}
\begin{center}
\hfill DTP-94/54 \\
\hfill September, 1994 \\
\vspace*{2cm}
{\LARGE \bf The Chiral Lagrangian parameters, $\overline{\ell}_1$,
$\overline{\ell}_2$,}
\vspace*{1cm} \\
{\LARGE \bf are determined by the $\rho$--resonance}
\vspace*{3cm} \\
{\large  \bf M.R.~Pennington, $\;$ J. Portol\'es}
\vspace*{0.5cm} \\
Center for Particle Theory , University of Durham \\
Durham  DH1 3LE,  U.K.
\vspace*{2.5cm} \\
\begin{abstract}
The all--important consequence of Chiral Dynamics for $\pi \pi$ scattering
is the Adler zero, which forces $\pi \pi$ amplitudes to grow asymptotically.
The continuation of this subthreshold zero into the physical regions
requires a $P$--wave resonance, to be identified with the $\rho$. It is
a feature of $\pi \pi$ scattering that convergent dispersive integrals
for the $I=1$ channel are essentially saturated by the $\rho$--resonance
and are much larger than those with $I=2$ quantum numbers. These facts
predict the parameters $\overline{\ell}_1$, $\overline{\ell}_2$ of the
Gasser--Leutwyler Chiral Lagrangian, as well as reproducing the well--known
KSFR relation and self-consistently generating the $\rho-$resonance.
\end{abstract}
\end{center}
\end{titlepage}
\newpage
\pagestyle{plain}
\pagenumbering{arabic}
\section{Introduction}
\hspace*{0.5cm} Interest in low energy pion dynamics has been rekindled
by developments in Chiral Perturbation Theory ($\chi$PT)
\cite{WE79,GL84,GL85} in the past decade.
 At lowest order in $\chi$PT, pion amplitudes are determined by just two
constants~: the pion decay constant, $f_{\pi}$, and the scale of explicit
chiral symmetry breaking which is, of course, set by the pion mass, $m_{\pi}$.
At the next order, new parameters, $\overline{\ell}_i \; \; (i=1,4)$ enter.
These have been fixed by Gasser and Leutwyler \cite{GL84}  by appeal to
detailed phenomenology. However, low energy hadron processes are for
the most part dominated by resonances. Thus, low energy $\pi \pi$
dynamics is determined by the $P$--wave $\rho$--resonance and by a strong
$S$--wave interaction, often called the $\sigma$ or $\epsilon$,
now the $f_{\circ}(1300)$. Of course,
Chiral Dynamics and resonance contributions are not in contradiction.
Indeed, the crucial link between these is provided by the continuity
of zero contours, Fig. 1. The Adler zero \cite{AD65} that controls
near threshold
$\pi \pi$ scattering becomes the Legendre zero of the $\rho$ \cite{PP712}, that
ensures that it is a spin--one resonance. The dominance of the $I=1$
$\pi \pi$ cross--section by the $\rho$--resonance at low energies and
the weakness of $I=2$ $\pi \pi$ interaction means that the only
parameters  needed to determine low energy $\pi \pi$ physics
are the mass and width of the $\rho$.
\par
It has long been known that the pion decay constant and the
$\rho$--parameters are connected by the KSFR relation \cite{KS66}.
Here, using dispersion relations and the continuity of
zero contours we show that the same $\rho$--parameters
not only provide relationships with the $\overline{\ell}_1$,
$\overline{\ell}_2$
of Gasser and Leutwyler as previously known \cite{GL84,EG89,DR89}, but
lead to a self-consistent
generation of the $\rho-$resonance.

\section{The high--low connection}

\hspace*{0.5cm} Chiral Dynamics is often thought of as imposing
constraints on low momentum processes and so attention is focussed on
these in isolation. However, our purpose is to emphasize that there is
a unity in hadron reactions that means that Chiral Dynamics affects
even high energy behaviour. Since this is germane to our discussion,
we begin by recalling this, allowing us to set up the relevant machinery.
\par
Consider the amplitude, $F(s,t)$, for $\pi^- \, \pi^{\circ} \rightarrow
\pi^- \, \pi^{\circ}$ in the $s$--channel. Chiral Dynamics imposes an Adler
zero at low energies in the scattering amplitude. This is, in fact, a
line of zeros through the Mandelstam triangle, Fig. 1, here obtained from
${\cal O}(p^4)$ $\chi$PT \cite{GL84} but experimental data
give a very similar contour \cite{PP712}. This line imposes
zeros in both the $s$ and $t$ channel $S$--wave amplitudes typically at
$s \simeq  m_{\pi}^2 / 2$ and $t \simeq m_{\pi}^2$, respectively.
This line means, for example, that at some fixed values of $t$ the
amplitude, $F(s,t)$, which is real within the Mandelstam triangle, changes
sign as $s$ increases, Fig. 1.
\par
Now this amplitude is known to have the appropriate cut plane analyticity
to satisfy fixed--$t$ dispersion relations for a range of $t$, in particular
for $0 \leq t \leq 4 m_{\pi}^2$. Let us first assume the asymptotic
behaviour of the amplitude is such that for $s \rightarrow \infty$, with
$t \in [ 0 , 4 m_{\pi}^2 ]$, $|F(s,t)| < const$. We can then write
an unsubtracted dispersion relation for this $s-u$ symmetric amplitude,
viz.
\begin{equation}
F(s,t) \; = \; \Frac{1}{\pi} \, \int_{4 m_{\pi}^2}^{\infty} \, ds' \,
{\cal I}m F(s',t) \, \left( \Frac{1}{s' - s} + \Frac{1}{s' - u} \right) \; \;
\; \; \; \; .
\label{eq:fstw}
\end{equation}
Since this amplitude describes the physical processes
$\pi^{\pm} \pi^{\circ} \rightarrow \pi^{\pm} \pi^{\circ}$ in the
$s$ and $u$--channels, it has a positive imaginary part for
$s' \geq 4 m_{\pi}^2$, $ t \in [ 0 , 4 m_{\pi}^2 ]$. Thus the
dispersive integral is positive definite for $ 0 \leq s,t,u \leq 4 m_{\pi}^2$
and the amplitude cannot have a zero in the Mandelstam triangle as Adler
requires.
\par
Consequently, Chiral Dynamics requires the amplitude $F(s,t)$ must grow
asymptotically, so that its dispersive representation must have
subtractions. Rigorously we know~\cite{MC70} $|F(s,t)| < s^2 $ as
$ s \rightarrow \infty$ for $ t \in [ 0 , 4 m_{\pi}^2 ]$, then once again
making use of the $s-u$ symmetry of $F(s,t)$, we have~:
\begin{equation}
F(s,t) \, = \, a(t) \, + \, \Frac{(s-u)^2}{4 \pi}
            \int_{4 m_{\pi}^2}^{\infty}  \Frac{ds'}{(s'
            - 2 m_{\pi}^2 + t/2 )^2}  \left( \Frac{1}{s' - s} +
            \Frac{1}{s' - u} \right) \; {\cal I}m F(s',t)
\label{eq:fstsub}
\end{equation}
where
\begin{equation}
a(t) \, = \, F ( s = u  \, , \, t \, ) \, \, \, \, \,
\, .
\end{equation}
Now we see that for those values of $t$ for which $a(t)$ (the amplitude
on the line $s=u$) is negative the amplitude will have a zero as $s$
increases. Thus asymptotically growing amplitudes may be regarded
as a consequence of, or at the very least are consistent with,
Chiral Dynamics.

\section{Calculating $\overline{\ell}_1$, $\overline{\ell}_2$}

\hspace*{0.5cm} Let $F^{x I}$ denote the $\pi \pi$ amplitude with
isospin $I$ in channel $x$, where $x = s$ or $t$. These amplitudes
can be decomposed into partial waves, $f_{\ell}^I(t)$, in the $t$-channel,
for example,
\begin{equation}
F^{tI} (t,s) \, = \,  \sum_{l=0}^{\infty} \, (2\ell +1) \, f_{\ell}^I(t) \,
                  P_{\ell} \left( 1 + \Frac{2s}{t - 4 m_{\pi}^2} \right)
                  \; \; \; \; .
\label{eq:exp}
\end{equation}
The scattering lengths $a_{\ell}^I$ are defined by the threshold limit~:
\begin{equation}
a_{\ell}^I \, = \, \lim_{t \rightarrow 4 m_{\pi}^2} \Frac{f_{\ell}^I(t)}{\left(
              \Frac{t}{4} - m_{\pi}^2 \right)^{\ell}} \; \; \; \; \; \; .
\label{eq:scatle}
\end{equation}
Our starting point is the fixed-$t$ dispersion relation for each isospin
amplitude, which rigorously needs no more than two subtractions
for $t \leq 4 m_{\pi}^2$. Partial wave projecting these and taking
$t \rightarrow 4 m_{\pi}^2$, we find the
$D$--wave scattering lengths given by the Froissart--Gribov representation
\begin{equation}
a_2^I \, = \, \Frac{16}{15 \pi} \, \int_{4 m_{\pi}^2}^{\infty} \,
\Frac{ds'}{s'^{3}} \, {\cal I}m F^I (s' , 4 m_{\pi}^2)
\label{eq:dwsl}
\end{equation}
for $I=0,2$.
This representation for the $D$--wave scattering lengths makes it clear
that the isospin combinations, for which the integral over the
absorptive parts, ${\cal I}m F ( s', 4 m_{\pi}^2)$, is positive, must
themselves be positive. Thus, we have
\begin{equation}
a_2^{0} - a_2^2 \geq 0 \; \; \; \; \; , \; \; \; \; \; \; \;
a_2^{0} + 2 a_2^2 \geq 0 \; \; \; \; \; .
\label{eq:ineq}
\end{equation}
In terms of ${\cal O}(p^4)$  $\chi$PT these scattering lengths are
combinations of the constants $\overline{\ell}_1$ and $\overline{\ell}_2$,
that appear in the Gasser--Leutwyler Lagrangian \cite{GL84}~:
\begin{eqnarray}
a_2^{0} & = & \Frac{1}{1440 \pi^3 F^4} \; \Big[ \; \overline{\ell}_1 \; +
\; 4 \overline{\ell}_2 \; - \; \Frac{53}{8} \; \Big] \; \; \; ,\nonumber \\
\label{eq:aa0222}
& & \\
a_2^2 & = &  \Frac{1}{1440 \pi^3 F^4} \; \Big[ \; \overline{\ell}_1 \; + \;
\overline{\ell}_2 \; - \; \Frac{103}{40} \;  \Big] \; \; \; .\nonumber
\end{eqnarray}
Then Eq. (\ref{eq:ineq}) leads to the straightforward constraints~:
\begin{equation}
\overline{\ell}_1 \; + 2 \overline{\ell}_2 \geq \Frac{157}{40} \; \; \; \; \;
\;
, \; \; \; \; \; \; \; \; \overline{\ell}_2 \geq \Frac{27}{20} \; \; \; .
\label{eq:nel12}
\end{equation}
These are displayed in Fig. 2 together with the previous phenomenological
values
\begin{equation}
 \overline{\ell}_1 = -2.3 \pm 3.7 \; \; \; \; \; , \; \; \; \; \; \; \;
 \overline{\ell}_2 = 6.0 \pm 1.3 \; \; \; \; ,
\label{eq:gasleut}
\end{equation}
determined by  Gasser and  Leutwyler \cite{GL84} using $a_2^{0}$ ,
$a_2^2$
of \cite{2a}. The simple
bounds, Eq.~(\ref{eq:nel12}), are close to the limits presented by
Wanders \cite{WA94} from the far more complicated conditions on the shape
of the subthreshold $\pi^{\circ} \pi^{\circ}$  $S$--wave that rigorously
also follow from $three$--channel crossing symmetry and positivity.
\par
Because of the explicit $1 / s'^{3}$ factor, the
integrals in Eq. (\ref{eq:dwsl}) are dominated by the low energy region,
since they converge
rapidly. Now each isospin amplitude, $F^{t I}$ $(I=0,1,2)$, can be
re--expressed in terms of $F^{s 1}$, $F^{s 2}$ and $F^{t 2}$ by the
isospin crossing matrix, for instance
\begin{equation}
F^{t 0} \, = \, \Frac{3}{2} \, F^{s 1} \,  +  \, \Frac{3}{2} \, F^{s 2}
               \,  +  \, F^{t 2} \; \; \; \; \; \; \; .
\label{eq:cross}
\end{equation}
Thus
\begin{eqnarray}
a_2^0 \, & = & \, \Frac{16}{15 \pi} \, \int_{4 m_{\pi}^2}^{\infty} \,
              \Frac{ds'}{s'^3} \, \bigg[ \, \Frac{3}{2} \, {\cal I}m
               F^{s 1} (s',4 m_{\pi}^2) \, + \, \Frac{3}{2} \, {\cal I}m
               F^{s 2} (s',4 m_{\pi}^2) \,  \nonumber \\
& \, & \\
\label{eq:a02im}
& &
\; \; \; \; \; \; \; \; \; \; \; \; \; \; \; \; \; \; \; \; \; \;
 \; \; +  \, {\cal I}m F^{t 2} (s', 4 m_{\pi}^2) \, \bigg]
\; \; \; \; \; \; \; \; . \nonumber
\end{eqnarray}
We then use the weakness of $I=2$ channels for physical mass pions to mean
\begin{equation}
\int_{4 m_{\pi}^2}^{\infty} \Frac{ds'}{s'^3}\; \left( {\cal I}m  F^{s 2} ,
                            {\cal I}m  F^{t 2} \right) \ll
\int_{4 m_{\pi}^2}^{\infty} \Frac{ds'}{s'^3} \;{\cal I}m  F^{s 1}
\label{eq:ineq1}
\end{equation}
so that
\begin{equation}
a_2^0  \, \simeq \, \Frac{16}{15 \pi} \,
              \int_{4 m_{\pi}^2}^{\infty} \,
            \Frac{ds'}{s'^3} \, \Frac{3}{2}\, {\cal I}m
            F^{s 1} (s', 4 m_{\pi}^2)
\; \; \; \; \; \; \; \; .
\label{eq:aproxdw}
\end{equation}
In the real world with physical pion mass
 this integral is dominated by the $\rho$--contribution, which can be
reliably evaluated in the narrow resonance approximation by
\begin{equation}
{\cal I}m F^{s 1} (s,t) \, = \, 3 \pi \, \sqrt{\Frac{s}{s - 4 m_{\pi}^2}} \, \;
                         m_{\rho} \Gamma_{\rho}  \, \left( 1 +
                         \Frac{2 t}{s - 4 m_{\pi}^2} \right)
                         \, \delta (s - m_{\rho}^2)
\label{eq:narrowi}
\end{equation}
so that
\begin{equation}
a_2^0 \, \Big|_{\rho} \;  = \, \Frac{24}{5} \, \Frac{\Gamma_{\rho}
             ( m_{\rho}^2 +
             4 m_{\pi}^2)}{m_{\rho}^4 ( m_{\rho}^2 - 4 m_{\pi}^2 )^{3/2}}
             \; \; \; \; \; \; \; .
\label{eq:dwnwi}
\end{equation}
For this approximation to make sense, clearly $|a_2^2| \ll a_2^0$.
We will check the consistency of this later.

While we are primarily concerned with the real world with physical mass pions,
it will be useful to compare our results with those of $\chi$PT.  Consequently,
we need to discuss what happens when the pion mass goes to zero.
To distinguish between the physical pion mass and a variable mass, we
denote the former by $M_{\pi}$ and the latter by $m_{\pi}$.  The inequalities
of Eq. (\ref{eq:ineq1}), of course, hold for physical pions ---each side
is logarithmically divergent when $m_{\pi} \to 0$.  We must therefore consider
the dispersive integral from $s^{\prime} = 4m_{\pi}^2$ to $4M_{\pi}^2$
separately.  This threshold contribution is readily evaluated using Weinberg's
 model amplitude \cite{Weinberg},
so that
\begin{equation}
{\cal I}m f^I_{\ell} (s) \, \simeq \, \sqrt{{s-4m_{\pi}^2}\over s}\,
({\cal R}e f^I_{\ell} (s))^2\; \; \; \; \; ,
\label{eq:imag}
\end{equation}
with
\begin{eqnarray}
{\cal R}e f^0_0(s)\,=\, {3\over 2} \,a^1_1\,(s - m_{\pi}^2/2) \; \; &
\, , \, & \; \;
{\cal R}e f^1_1(s)\,=\, {1\over 4} \,a^1_1\,(s-4m_{\pi}^2) \; ,  \nonumber \\
& & \\
\label{eq:weinb}
{\cal R}e f^2_0(s) & = &  {3\over 4} \,a^1_1\,(2m_{\pi}^2-s)\quad , \nonumber
\end{eqnarray}
where $a^1_1$ is the $P$-wave scattering length.  Then we easily deduce that
the \lq\lq chiral" components of $a^0_2$ and $a^2_2$ are respectively~:
\begin{eqnarray}
a^0_2 \Big|_{\chi} & = & {2\over\pi}\,(a^1_1)^2\,\left[\ln
\left({{1+X}\over{1-X}}
\right) - 2X - {91 X^3\over{240}} - {X^5\over{80}}\right] \; \; , \nonumber \\
\label{eq:X}
& & \\
 a^2_2 \Big|_{\chi} & = & {4\over{5\pi}}\,(a^1_1)^2\,
\left[\ln\left({{1+X}\over{1-X}}
\right) - 2X - {13 X^3\over{96}} - {11 X^5\over{160}}\right] \; \; ,\nonumber
\end{eqnarray}
where $X^2 = 1 - m_{\pi}^2/M_{\pi}^2$.
These threshold contributions are to be added to the $\rho-$component
of Eq. (\ref{eq:dwnwi}), for example.  However, for physical mass pions, i.e.
with
$m_{\pi} =M_{\pi}$, $X=0$, these \lq\lq chiral" components vanish.
\par
Because the intercept of the $\rho$--Regge trajectory is below one, the
$I=1$ $t$--channel amplitude divided by $(s-u)$ satisfies an
unsubtracted dispersion relation
for $t \leq 4 m_{\pi}^2$. Projecting out the $P$--wave and expanding for
$t \simeq 4 m_{\pi}^2$ gives
\begin{eqnarray}
\Frac{f_1^1(t)}{\Big( \Frac{t}{4} - m_{\pi}^2 \Big) } \, &  = &  \,
\Frac{4}{3 \pi} \, \int_{4 m_{\pi}^2}^{\infty} \, ds' \, \left[
\Frac{1}{s'^2} - \Frac{(t - 4 m_{\pi}^2)}{s'^3} + \ldots \right] {\cal I}m
F^{t 1}(s',t) \nonumber \\
\label{eq:thresh}
& & \\
& \simeq & a_1^1 \, + \, b_1^1 \left( \Frac{t}{4} - m_{\pi}^2 \right) \, +
\ldots \nonumber
\end{eqnarray}
then
\begin{eqnarray}
a_1^1 & = & \Frac{4}{3 \pi} \, \int_{4 m_{\pi}^2}^{\infty} \,
\Frac{ds'}{s'^2} \, {\cal I}m F^{t 1} (s' , 4 m_{\pi}^2) \label{eq:a11} \\
& & \nonumber \\
b_1^1 & = & \Frac{16}{3 \pi} \, \int_{4 m_{\pi}^2}^{\infty} \,
\Frac{ds'}{s'^2} \, \left[ \Frac{\partial}{\partial t} {\cal I}m
F^{t 1}(s' , 4 m_{\pi}^2) \, - \, \Frac{1}{s'} \, {\cal I}m
F^{t 1}(s' , 4 m_{\pi}^2) \, \right] \; \; \; \; \; \; \; \; \; \; .
\label{eq:b11}
\end{eqnarray}
Note that integral of Eq. (\ref{eq:a11}) has an explicit factor of
$1/s'^2$, while that of Eq. (\ref{eq:b11}), like Eq. (\ref{eq:dwsl}),
has $1/s'^3$. Consequently, the integral for $a_1^1$ is not so dominated
by the low energy $\rho$--contribution. Nevertheless, this
$\rho$--contribution gives, using $F^{t 1} = F^{s 1} - F^{s 2} + F^{t 2}$
\begin{equation}
a_1^1 \, \Big|_{\rho} \,   = \, \Frac{4 \Gamma_{\rho} (m_{\rho}^2 +
                    4 m_{\pi}^2 )}{m_{\rho}^2 ( m_{\rho}^2 -
                   4 m_{\pi}^2)^{3/2}} \; \; \; ,
\label{eq:a11rho}
\end{equation}
\begin{equation}
b_1^1 \, \Big|_{\rho} \,  =  \, \Frac{16 \Gamma_{\rho}}{m_{\rho}^4
                      ( m_{\rho}^2 -
                   4 m_{\pi}^2)^{1/2}} \; \; \; \; \; \; \; \; .
\label{eq:b11rho}
\end{equation}
In the chiral limit $f^1_1(t)$ displays no logarithms.  To show that the
effective range $b_1^1$ also has no logarithms of $m_{\pi}^2$, care must
be taken in the order of limits $m_{\pi} \to 0$ and  $t \to 4 m_{\pi}^2$.
\par
Now in the chiral limit when $m_{\pi} \to 0$,
the $\rho$--contributions give~:
\begin{eqnarray}
a_0^0,a_0^2 \rightarrow 0 & \; \; , \; \; \; &
a_1^1 \, \Big|_{\rho} \, \rightarrow \Frac{4 \Gamma_{\rho}}{m_{\rho}^3}
\; \; \; \; , \nonumber \\
\label{eq:chlim}
& & \\
b_1^1 \, \Big|_{\rho} \, \rightarrow \Frac{16 \Gamma_{\rho}}{m_{\rho}^5} &
\; \; \; , \; \; \;
\; & a_2^0 \, \Big|_{\rho} \, \rightarrow \Frac{24}{5}
                \Frac{\Gamma_{\rho}}{m_{\rho}^5}
\; \; \; \; \; \; \; , \nonumber
\end{eqnarray}
while the near threshold contribution of Eq.~(\ref{eq:X}) yields
\begin{equation}
a^0_2 \Big|_{\chi} \; = \; {2\over\pi}\,(a^1_1)^2\,
\ln\left({4M_{\pi}^2\over{
m_{\pi}^2}}\right) \; \; \; \;  \; \; , \; \; \; \; \;
a^2_2 \Big|_{\chi} \;  = \;  {4\over{5\pi}}\,(a^1_1)^2\,
\ln\left({4M_{\pi}^2\over{
m_{\pi}^2}}\right) \; \; \; ,
\end{equation}
when $m_{\pi}^2 \ll M_{\pi}^2$.
In $\chi$PT in the same limit
\begin{eqnarray}
a_1^1 & \rightarrow & \Frac{1}{24 \pi F^2} \; \; \; \; \; , \nonumber \\
b_1^1 & \rightarrow & \Frac{1}{288 \pi^3 F^4} \, \left[ \, - \overline{\ell}_1
\, + \, \overline{\ell}_2 \, + \, \Frac{97}{120} \, \right] \; \; \; ,
\label{eq:chptlim}
\\
a_2^0 & \rightarrow & \Frac{1}{1440 \pi^3 F^4} \, \left[ \,
\overline{\ell}_1 \, + \, 4 \, \overline{\ell}_2 \, - \, \Frac{53}{8} \,
\right]
\;\;\; . \nonumber
\end{eqnarray}

Simply equating and combining these results
gives a straightforward  idea of the size
of $\overline{\ell}_1, \overline{\ell}_2$. We find
\begin{eqnarray}
\Frac{1}{96 \pi F^2} & \, = \, & \, \Frac{\Gamma_{\rho}}{m_{\rho}^3} \; \; \; ,
\nonumber
\\
\label{eq:rcomp}
\overline{\ell}_1 & \, = \, & \, \Frac{1183}{600} \, - \, \Frac{\pi}{4}
 \Frac{m_{\rho}}{\Gamma_{\rho}} \; + \; \overline{\ell}_1 \Big|_{\chi} \;
\; , \\
\overline{\ell}_2 & \, = \, & \, \Frac{349}{300} \, + \, \Frac{\pi}{4}
\Frac{m_{\rho}}{\Gamma_{\rho}} \;+\; \overline{\ell}_2 \Big|_{\chi} \; \; \;
 \; ,\nonumber
\end{eqnarray}
where
\begin{equation}
 \overline{\ell}_1 \Big|_{\chi}\, = \,\ln\left({4M_{\pi}^2\over{m_{\pi}^2}}
\right) \;\; \; \; \; \; \; , \; \; \; \; \; \; \; \;
 \overline{\ell}_2 \Big|_{\chi}\, = \,\ln\left({4M_{\pi}^2\over{m_{\pi}^2}}
\right) \; \;  .
\label{eq:lchi}
 \end{equation}
--- in Eq. (\ref{eq:lchi}), we drop the constants implied
by Eq. (\ref{eq:X}), since if $m_{\pi}^2 \ll M_{\pi}^2$, these are
irrelevant.\par
 The first of the relations in Eq.~(\ref{eq:rcomp}) yields the long
established KSFR relation \cite{KS66}.
Moreover, Eqs.~(\ref{eq:rcomp}, \ref{eq:lchi})  reproduce the chiral logs
of $\chi$PT \cite{GL84},
 but with their renormalization scale $\mu$ fixed by Eq. ({\ref{eq:ineq1}}).
 Of course, for the real world $\overline{\ell}_1 \Big|_{\chi} =
  \overline{\ell}_2 \Big|_{\chi} \,=\,0$ (cf. Eq.~(\ref{eq:X}) with
$X=0$) and then these
  simplified relations give with the physical $\rho$ mass and
  width \cite{PD94}
\begin{equation}
F \, = \, 99.6 \, {\rm{MeV}} \, \, , \; \; \; \; \; \;
 \overline{\ell}_1 \, = \, -2.01 \; \; , \; \; \; \; \; \;
 \overline{\ell}_2 \, = \, 5.15
\label{eq:fres}
\end{equation}
and
\begin{equation}
\Frac{a_2^2}{a_2^0} \; = \; \Frac{56 \Gamma_{\rho}}{75 \pi m_{\rho}} \;
= \; 0.047
\label{eq:relief}
\end{equation}
which is reassuringly small.
\par
To have an idea of how certain these numbers are, we could, at the
same level of approximation, regard Eq. (\ref{eq:aproxdw}) as a
calculation of $a_2^0 - a_2^2$ rather than $a_2^0$. Then with
$\overline{\ell}_1 \Big|_{\chi} \, = \, \overline{\ell}_2 \Big|_{\chi} \,
= \, 0$
\begin{eqnarray}
\overline{\ell}_1 \, & = & \, \Frac{259}{120} \, - \, \Frac{\pi}{4}
\Frac{m_{\rho}}{\Gamma_{\rho}} \,  = \; -1.82 \; \; \; \; ,\nonumber \\
\label{eq:otherw}
& & \\
\overline{\ell}_2 \, & = & \, \Frac{27}{20} \, + \, \Frac{\pi}{4}
\Frac{m_{\rho}}{\Gamma_{\rho}} \, = \; 5.33 \nonumber
\end{eqnarray}
and
\begin{equation}
\Frac{a_2^2}{a_2^0} \, = \, 0.072 \; \; \; \; \; .
\label{eq:othratio}
\end{equation}
\par
We now turn to the full results with physical pion mass. Eqs.
(\ref{eq:aproxdw},\ref{eq:b11}) are dominated by the $\rho$--resonance
to an accuracy of
better than $20 \%$. To take into account this uncertainty we introduce
a factor $\lambda = 1.0 \pm 0.2$. Then, with $f_{\pi} = (90 \pm 2) \,
\rm{MeV}$ and  $m_{\pi} = 138 \, {\rm MeV}$ (both the mean of their
charged and neutral values), we have  from Eqs. (\ref{eq:dwnwi},
{}~\ref{eq:b11rho},~\ref{eq:chptlim})
\begin{eqnarray}
\overline{\ell}_1 \, + \, 4 \overline{\ell}_2 \, - \, \Frac{53}{8} \; & = &
\; 6912 \pi^3 \, \Frac{\Gamma_{\rho} f_{\pi}^4}{m_{\rho}^5} \,
\Frac{\left( \, 1 \, + \, \Frac{4 m_{\pi}^2}{m_{\rho}^2} \, \right)}{ \left(
\, 1 \, - \, \Frac{4 m_{\pi}^2}{m_{\rho}^2} \, \right)^{3/2}} \,
\lambda \nonumber \\
\label{eq:comb1}
& & \\
& = & 10.9 \pm 2.4 \nonumber
\end{eqnarray}
and
\begin{eqnarray}
- \overline{\ell}_1 \, + \, \overline{\ell}_2 \, + \, \Frac{97}{120}  \; &
= & \;
4608 \pi^3 \, \Frac{\Gamma_{\rho} f_{\pi}^4}{m_{\rho}^5} \,
\Frac{\lambda}{\left(\, 1 \, - \, \Frac{4 m_{\pi}^2}{m_{\rho}^2} \,
\right)^{1/2}} \nonumber \\
\label{eq:comb2}
& & \\
& = & 5.6 \pm 1.2 \; \; \; \; \; \; \; \; \; .\nonumber
\end{eqnarray}
These yield our main result~:
\begin{equation}
\overline{\ell}_1  \; = \; -0.3 \, \pm \, 1.1 \; \; \; , \; \; \; \; \;
\overline{\ell}_2  \; = \; 4.5 \, \pm \, 0.5 \; \; \; \; \; \; \; \; \; \; ,
\label{eq:mainres}
\end{equation}
which are in agreement with the values deduced by Gasser and Leutwyler
\cite{GL84}  given in Eq. (\ref{eq:gasleut}),
but with considerably reduced errors, which in Eqs.
(\ref{eq:comb1},\ref{eq:comb2}) remain on the conservative side. These
results are shown in Fig. 2 together with the determinations by
Riggenbach et al. from a fit of calculations in $\chi$PT with experimental
$\pi \pi$ and $K_{l4}$ parameters \cite{RG91} and with the estimate
by Beldjoudi and Truong \cite{BT94} found by fitting the $\pi \pi$ $P$--wave
and the isoscalar $S$--wave with one loop $\chi$PT unitarized by
$[1,1]$ Pad\'e approximant \cite{BT94,DH90}.
\par
An explanation of why we do not use the Froissart--Gribov representation
of the $P$--wave scattering length, Eq.~(\ref{eq:a11}), to determine
the $\overline{\ell}_i$ is in order. Despite the fact that $\rho$--dominance
of the integral of Eq.~(\ref{eq:a11}) does lead to the KSFR relation
(the first relation in Eq.~(\ref{eq:rcomp})) in the chiral limit,
nonetheless, the sum rule of Eq.~(\ref{eq:a11}) does have two sources
of appreciable corrections both resulting from its slower convergence
than the integrals of Eqs.~(\ref{eq:ineq1},\ref{eq:b11})~: one is the
high energy contribution, the other are corrections to the assumed
weakness of $I=2$ components (i.e. the analogue of Eq.~(\ref{eq:ineq1})
with $s'^2$ in the denominator). It seems these corrections unexpectedly
cancel. The $\rho$--contribution of Eq.~(\ref{eq:a11rho}) gives
$a_1^1 \, |_{\rho} \, = 0.035$. With the values of $\overline{\ell}_1,
\overline{\ell}_2$ we have obtained substituted into Eq.~(18.5) of Ref. 2,
${\cal O} (p^4)$ $\chi$PT gives $a_1^1 \, |_{\chi PT} \, \, = \; 0.036
\pm 0.002$ --- a very similar value.
However, because of the corrections to Eq.~(\ref{eq:a11rho}), $a_1^1$
provides a consistency check rather than a vehicle for determining the
$\overline{\ell}_i$.\par
In $\chi$PT it is the chiral logarithms, typically with  a scale of
$\mu = m_{\pi}$ \cite{GL84}, that give
the major contribution to the $D$-wave scattering length $a^0_2$, for instance,
while the explicit $\rho-$resonance component is smaller.  In contrast, here
the \lq\lq chiral logarithms" play no
role when the pion mass is 138 MeV, see Eq. (\ref{eq:X}), and the
whole answer, Eqs. (\ref{eq:aa0222}, \ref{eq:comb1}) is given
by the $\rho-$component.  This is a direct consequence of the physical
assumption that Eq.~(\ref{eq:ineq1}) holds for $m_{\pi}\ = M_{\pi}$.\par
That $\overline{\ell}_1, \overline{\ell}_2$ are directly relatable
to the $\rho-$resonance has already  been considered in \cite{GL84,DR89}
using vector meson dominance.  Their idea is to couple the $\rho-$meson
to the pion
in a chirally symmetric way and then to assume that elastic $\pi\pi$
scattering is dominated by
$\rho-$exchange.  By comparing the effective Lagrangian obtained in the limit
of $p^2 \ll m_{\rho}^2$ (i.e. neglecting the momentum dependence in the
$\rho-$propagator)
with the $SU(2) \times SU(2)$ $\chi$PT Lagrangian at $O(p^4)$, Gasser and
Leutwyler are able
to determine the $\rho-$contribution to $\overline{\ell}_1$,
$\overline{\ell}_2$,
that happens to saturate the phenomenologically determined values.
 An extension of this procedure to
$SU(3) \times SU(3)$ including the lightest (non-Goldstone) meson
spectrum~: scalars, pseudoscalars, vectors and axial vectors, has been studied
in Ref. \cite{EG89}.
\par
As we have stressed
above, our procedure is also based on $\rho-$dominance of the
$I=J=1$ $\pi\pi$
channel, together with the  well established phenomenological fact that
exotic ($I=2$) channels have comparatively small absorptive parts.
In the next section we shall show that the
zero contours of the $\pi\pi$ amplitude allow us to check
the consistency of these twin assumptions.

\section{Zeros connected}
\baselineskip=6.95mm

\hspace*{0.5cm} As already emphasized, a key feature of $\pi \pi
\rightarrow \pi \pi$ scattering is the appearance of the Adler zero
on--shell \cite{PP712}. Moreover, amplitudes being analytic functions of
several complex
variables, this zero is not isolated but lies on a line that passes
through the Mandelstam triangle, Fig. 1. This zero may (depending on the
channel) continue into the scattering regions and thereby generate a dip
in the angular distribution that can be measured. Thus, if we consider
the amplitude for $\pi^+ \pi^- \rightarrow \pi^{\circ} \pi^{\circ}$
in the $t$--channel, the zero enters the $s$ and $u$--channel
$\pi^- \pi^{\circ} \rightarrow \pi^- \pi^{\circ}$ physical regions. That
this zero contour curves down, Fig. 1, is a consequence of the $D$--wave
scattering length, $a_2^{0} - a_2^2$ being positive, Eq. (\ref{eq:ineq}).
Now, well into the $s$ and $u$--channels, in the $\rho$--region, if the
amplitude is dominated by the $P$--wave, the angular distribution would have
a marked dip at $\cos \theta = 0$, reflecting the spin--one nature of
the $\rho$--resonance~:
\begin{equation}
F (\pi^- \pi^{\circ} \rightarrow \pi^- \pi^{\circ}) \simeq
\Frac{3}{2} \; \, \Frac{m_{\rho} \Gamma (s)}{m_{\rho}^2 - s - i m_{\rho}
\Gamma(s)} \cos \theta_s + \ldots
\label{eq:rhho}
\end{equation}
In reality, the $\rho$ has an $S$--wave background, which means the
minimum in the differential cross--section is not exactly zero nor exactly
at $\cos \theta_s = 0$. Nevertheless, there is a dip and this
corresponds to the zero of the amplitude having moved into the complex
plane.
However, it does not move far away, since, in this channel, the $S$--wave
background has isospin two and so is small. This zero follows the track
shown in Fig. 1 for  ${\cal R}e$ $s$, ${\cal R}e$ $t$. Clearly it
connects to the Adler zero.
\par
In $\chi$PT, a resonance, like the $\rho$, is not generated at any finite
order and so the predictions of $\chi$PT are not realistically continuable
much above $\pi \pi$ threshold without some technique, typically
Pad\'e approximants \cite{DH90}, for estimating the all orders sum from
the known low order predictions. However, there is an alternative
procedure, which we present here, that provides a consistency check on the
values of $\overline{\ell}_1, \overline{\ell}_2$, we have just deduced from
the $\rho$--parameters.
\par
It is a feature of zeros of amplitudes that they generally continue
smoothly from one region to another \cite{???}, even if the corresponding
amplitudes are not well determined. Thus, we assume that though the one
loop prediction from $\chi$PT for the $\pi \pi$ amplitude is not to be
believed beyond $\sim 600$ MeV, the zeros of this amplitude are more
reliably given --- even up to $900$ MeV. In Fig. 1, we have shown the track of
the corresponding minimum of the differential cross--section in the $s$
and $u$ $\pi^- \pi^{\circ} \rightarrow \pi^- \pi^{\circ}$ channels whether
from ${\cal O}(p^4)$ $\chi$PT or experiment.
Taking the $I=2$ $S$--wave, through this region, to be that parameterized
by Schenk \cite{SC91} which matches on to $\chi$PT near threshold, one can,
knowing the zero, predict the corresponding behaviour of the $P$--wave.
Thus from the minimum at $\cos \theta_s = z (s)$, shown in Fig. 1, we
have, assuming elastic unitarity,
\begin{equation}
\tan \delta_1^1 (s) \; = \; \Frac{- \, \Frac{1}{2} \, \sin 2
\delta_{0}^2 (s) }{ 3 \, z(s) + \sin^2 \delta_{0}^2 (s)}
\label{eq:deltas}
\end{equation}
where $\delta_{0}^2$ and $\delta_1^1$ are the $I=2$ $S$--wave and
$I=1$ $P$--wave phase shifts. Note that since the imaginary part of
the $I=2$ amplitude is small, the $P$--wave phase $\delta_1^1 \rightarrow
\pi/2$, where $z(s) \rightarrow 0$, i.e. where the zero is exactly
in the middle of the angular distribution of Eq.~(\ref{eq:rhho}).
With Schenk's parameterization for $\delta_{0}^2$, we predict the
$P$--wave phase shown in Fig.~3 , using the zero determined by one
loop $\chi$PT with $\overline{\ell}_1 = -0.3$, $\overline{\ell}_2 = 5.0$,
within the range we expect, Eq. (\ref{eq:mainres}). This prediction
is compared with the LBL \cite{LBL} and CERN-Munich phase--shifts, the
latter from both the analysis by Ochs \cite{OC73} and that by Estabrooks
and Martin \cite{EM74}. We also display in the same Fig. 3 the prediction
of ${\cal O}(p^4)$ $\chi$PT using $\delta_1^1 = \sqrt{1-4m_{\pi}^2/s} \,
{\cal R}e  f_1^1$ agreed as the correct interpretation of the chiral
prediction for $\delta_1^1$ from ${\cal R}e  f_1^1$ \cite{GM91,DP93}. The
contrast is marked.
This comparison illustrates
\begin{itemize}
\item[(i)] the consistency of our determination of $\overline{\ell}_1$,
$\overline{\ell}_2$, and
\item[(ii)] that though the $P$--wave in one loop $\chi$PT
is not reliable beyond $\sim 600$ MeV or so, the zero $\chi$PT predicts is much
less affected by as yet uncalculated higher order corrections.
Consequently, this provides a physically motivated unitarization procedure,
bringing good agreement with experiment.
\end{itemize}

\section{Conclusion}

\hspace*{0.5cm} $\rho$--dominance of the $I=1$ $\pi \pi$ cross--section
and the relative weakness of $I=2$ interactions leads to values for the
parameters of the $\chi$PT Lagrangian \cite{GL84} $\overline{\ell}_1 =
-0.3 \pm 1.1$, $\overline{\ell}_2 = 4.5 \pm 0.5$. These values give a
zero of the $\pi^- \pi^{\circ} \rightarrow \pi^- \pi^{\circ}$ amplitude
that continues from the Adler zero below threshold to the Legendre zero
of the $\rho$--resonance. Such a smooth continuation of the Adler zero
demands the existence of a spin--one resonance --- a resonance we know
as the $\rho$. Consequently, it is natural that the constants
$\overline{\ell}_1,
\overline{\ell}_2$ of the
Chiral Lagrangian should be fixed self-consistently by the
$\rho$--parameters, as we have shown.
\newpage

{\bf Acknowledgements}
\vspace*{0.5cm} \\
\hspace*{0.5cm} It is a pleasure to thank J\"urg Gasser for useful
discussions and correspondence. JP is grateful for the support of
grant ERBCMRXCT 920026 of the EC Human and
Capital Mobility
programme EuroDA$\Phi$NE network. JP is also partially supported by
 DGICYT PB091--0119 at the University of Valencia (Spain).

\vspace*{1cm}

\newpage
\hspace*{-0.6cm}{\Large \bf Figure Captions}
\\
\\
{\bf Fig. 1}
\vspace*{0.3cm} \\
\hspace*{0.5cm} The track of the zero in the amplitude for
             $\pi^- \pi^{\circ} \rightarrow \pi^- \pi^{\circ}$ (in the
             $s$--channel) in the Mandelstam plane obtained
             from ${\cal O}(p^4)$ $\chi$PT \cite{GL84}. Where the zero is at
             complex $s$ (in the physical regions), the ${\cal R}e \, s$ is
             plotted.
\\
\\
\\
{\bf Fig. 2}
\vspace*{0.3cm} \\
\hspace*{0.5cm} Predictions for $\overline{\ell}_1, \overline{\ell}_2$ of the
Gasser--Leutwyler Chiral Lagrangian. The dashed lines mark the {\it
positivity}  bounds of Eq.~(\ref{eq:nel12}). The shaded  ellipse defines
the present results. $\Diamond$ is the first evaluation by
Gasser and Leutwyler \cite{GL84}, $\odot$ is the updated fit
by Riggenbach et al. \cite{RG91} and $\Box$ the central value from
the fit by Beldjoudi and Truong \cite{BT94} for which no errors were
determined.
\\
\\
\\
{\bf Fig. 3}
\vspace*{0.3cm} \\
\hspace*{0.5cm} The $\pi \pi$ $I=1$, $P$--wave phase shift, $\delta_1^1$,
below $1 {\rm GeV}$. The solid line is the present prediction based on
the zero contour of Fig. 1, as described in the text. The dashed line is
the ${\cal O}(p^4)$ $\chi$PT result for $\sqrt{1-4 m_{\pi}^2/s} \, {\cal R}e
  f_1^1$,
which is the agreed definition of $\delta_1^1$  at this order \cite{GM91,DP93}.
The data are from the energy--independent analyses by Protopopescu et al.
\cite{LBL} ($\Box$) of the LBL experimental results, by Ochs \cite{OC73}
($\odot$) and by Estabrooks and Martin \cite{EM74} ($\bigtriangleup$)
of the CERN--Munich data.

\begin{thebibliography}{99}


\bibitem{WE79} S. Weinberg, {\em Physica} {\bf 96A} (1979) 327.


\bibitem{GL84} J. Gasser and H. Leutwyler, {\em Ann. Phys.}(NY)
                     {\bf 158} (1984) 142.

\bibitem{GL85} J. Gasser and H. Leutwyler, {\em Nucl. Phys.}
                     {\bf B250} (1985) 465.

\bibitem{AD65} S. Adler, {\em Phys. Rev.} {\bf 137B} (1965) 1022,
                     {\bf 139B} (1965) 1638.


\bibitem{PP712} M.R. Pennington and S. D. Protopopescu, {\em Phys. Lett.}
              {\bf B40} (1972) 105; \\
                M.R. Pennington, {\em Proc. of Conf. on $\pi \pi$ scattering},
                Tallahassee, 1973, AIP Conf. Proc. 13 (ed. P.K. Williams and
                V. Hagopian), p.89.

\bibitem{KS66} K. Kawarabayashi and M. Suzuki, {\em Phys. Rev. Lett.}
                     {\bf 16} 255 (1966) 255; \\
               Riazuddin and  Fayyazuddin, {\em Phys. Rev.}
                     {\bf 147} (1966) 1071.


\bibitem{EG89}  G. Ecker, J. Gasser, A. Pich and E. de Rafael, {\em Nucl.
Phys.}
                     {\bf B321} (1989) 311.

\bibitem{DR89} J.F. Donoghue, C. Ramirez and G. Valencia,  {\em Phys. Rev.}
                     {\bf D39} (1989) 1947.

\bibitem{MC70} A. Martin and F. Cheung, {\em ``Analytic Properties
                    and Bounds of the Scattering Amplitude"}, Gordon and
                    Breach, (1970).

\bibitem{2a} J.L. Petersen, CERN Yellow Report 77-04 (1977).


\bibitem{WA94} G. Wanders, talk at the Workshop on {\em Chiral Dynamics~:
               Theory and Experiment}, MIT, July 1994~;
               B. Ananthanarayan, D. Toublan and G. Wanders,
               Universit\'e de Lausanne preprint UNIL-TP-4/94, hep-ph/9410302
(October, 1994).

\bibitem{Weinberg} S. Weinberg, {\em Phys. Rev. Lett.} {\bf 17} (1966)
                   616.

\bibitem{PD94} Review of Particle Properties, {\em Phys. Rev.}
                     {\bf D50} (1994) 1173.


\bibitem{RG91} C. Riggenbach, J. Gasser, J.F. Donoghue, and B.R. Holstein,
               {\em Phys. Rev.} {\bf D43} (1991) 127.

\bibitem{BT94} L. Beldjoudi and T.N. Truong, \lq \lq $\pi \pi$ scattering
and pion form factors", \'Ecole Polytechnique preprint CPTH-A292.0294 (1994).



\bibitem{DH90} A. Dobado, M.J. Herrero and T.N. Truong, {\em Phys. Lett.}
                     {\bf B235} (1990) 134.



\bibitem{???} R. Odorico, {\em Phys. Lett.} {\bf B38} (1972) 411;\\
              E. Barrelet, {\em Il Nuovo Cimento} {\bf 8A} (1972) 331.

\bibitem{SC91} A. Schenk, {\em Nucl. Phys.} {\bf B363} (1991) 97.


\bibitem{LBL} S.D. Protopopescu et al., {\em Phys. Rev.} {\bf D7} (1973)
              1279.

\bibitem{OC73} W. Ochs, thesis submitted to the University of Munich (1973).

\bibitem{EM74} P. Estabrooks and A.D. Martin, {\em Nucl. Phys.} {\bf B79}
               (1974) 301.


\bibitem{GM91} J. Gasser and Ulf-G. Meissner, {\em Phys. Lett.} {\bf B258}
               (1991) 219.

\bibitem{DP93} A. Dobado and J.R. Pel\'aez, {\em Z. Phys.} {\bf C57} (1993)
501.

\end{thebibliography}
\end{document}